\documentclass[twocolumn,eqsecnum,preprintnumbers,superscriptaddress,prb,amsmath,amssymb,floatfix]{revtex4}
\usepackage{graphicx}
\usepackage{dcolumn}
\usepackage{bm}
\usepackage{subfigure}
\usepackage{hyperref}
\def\be{\begin{equation}}
\def\ee{\end{equation}}
\def\ba{\begin{eqnarray}}
\def\ea{\end{eqnarray}}

\def\BSCCO{Bi$_2$Sr$_2$CaCu$_2$O$_{8+\delta}$}
\def\C60{A$_x$C$_{60}$}

\def\HgCu3{HgCa$_2$Cu$_3$O$_{8+y}$}
\def\HgCu4{HgBa$_2$Ca$_3$Cu$_4$O$_{10+y}$}
\def\TlCu3{Tl$_2$Ba$_2$Ca$_2$Cu$_3$O$_{10+y}$}
\def\TlCu4{Tl$_2$Ba$_2$Ca$_3$Cu$_4$O$_{12+y}$}

\def\BiCu3{Bi$_2$Sr$_2$Ca$_{2}$Cu$_3$O$_y$}

\def\C60{A$_x$C$_{60}$}

\newcommand{\rr}{\mathbf{r}}

\newcommand{\kk}{\mathbf{k}}

\begin{document}

\title{Inferring effective interactions from the local density of states: application to STM data from \BSCCO}
\author{R.Jamei}
\author{J.Robertson}
\author{E-A.Kim}
\affiliation{Department of Physics, Stanford University, Stanford, California  94305}

\author{A.Fang}
\affiliation{Department of Applied Physics, Stanford University, Stanford, California  94305}

\author{A.Kapitulnik}
\affiliation{Department of Physics, Stanford University, Stanford, California  94305}
\affiliation{Department of Applied Physics, Stanford University, Stanford, California  94305}

\author{S.A.Kivelson}
\affiliation{Department of Physics, Stanford University, Stanford, California  94305}

\begin{abstract}
While the influence of impurities on the local density of states (LDOS) in a metal is notoriously non-local due to interference effects, low order moments of the LDOS in general can be shown to depend only on the local structure of the Hamiltonian.  Specifically, we show that an analysis of the spatial variations of these moments permits one to ``work backwards'' from  scanning tunneling microscopy (STM) data to infer the local structure of the underlying effective Hamiltonian. Applying this analysis to STM data from the high temperature superconductor, {\BSCCO}, we find that the variations of the electro-chemical potential are remarkably small ({\it i.e.,} the disorder is, in a sense, weak) but that there are large variations in  the local magnitude of the $d$-wave gap parameter.

\end{abstract}


\maketitle

\section{Introduction}
Scanning tunneling microscope (STM) measurements on highly correlated materials, such as the high temperature superconductors, are increasingly being analyzed \cite{scalapino, hirschfeld_1, kapitulnik_1, davis_m, nunner, miyakawa, yazdani, hoffman, davis_1,lang,cren,fang,kapitulnik_2,davis_2,davis_3, DHLee, balatsky, balatsky_2, chakravarty, john, sak, scalapino_2} in the hopes of inferring information concerning the character of  the dominant interactions affecting the electron dynamics.  However, it has long been known, as exemplified by a series of famous experiments~\cite{crommie_1,hari_1}
on simple metal surfaces, that the effects on the local density of states (LDOS) of a change in the Hamiltonian are highly {\it non-local}.  The interference patterns produced by quasiparticles scattering from an impurity cause ripples in the LDOS to spread far and near.  The possibility of such non-local effects make it a non-trivial issue how to best solve the inverse problem:  given a set of high quality STM data, how does one work backwards to infer the character of the effective Hamiltonian that produced the measured spectra?  In the present paper, we show that the low order moments of the LDOS are determined entirely by the local form of the Hamiltonian, and that they can therefore be simply analyzed to obtain an approximate solution to this inverse problem.  
 
Of particular interest to us are STM studies of the high temperature superconductor, {\BSCCO} (BSCCO), which have 
revealed\cite{scalapino,cren,davis_m,hoffman,kapitulnik_1,kapitulnik_2,fang,davis_1,lang} intriguing spatial inhomogeneities in the  LDOS, $\rho_\rr(\omega)$:  For energies $\omega$ of order the superconducting gap, $ \Delta_0$, there are order one variations in the LDOS, although for $|\omega| \ll \Delta_0$, and again for  $|\omega| \gg \Delta_0$ there are only relatively small fractional variations in $\rho$ as a function of position, $\rr$.   The energy of the low energy peak in $\rho_\rr(\omega)$ (usually called the superconducting coherence peak) varies substantially from one location
 to another, with a ``patch'' diameter of around 30$\AA$.  
There have been at least two different interpretations of  these results proposed:
The first proposed interpretation~\cite{davis_1,kapitulnik_2} associates the local value of the peak in $\rho_\rr(\omega)$
with the local magnitude of the gap parameter,  $\Delta_{\rr,\rr^\prime}$, in a mean field Hamiltonian.
 However, Fang~\emph{et al.}~\cite{scalapino} showed that this interpretation
can be quite misleading.  
They calculated the LDOS of a metallic region
(with a local value of $\Delta = 0 $) embedded in the background of a $d$-wave superconductor, with a bulk $\Delta = \Delta_0$.  Unsurprisingly, they found that for a
patch size comparable to the superconducting coherence length,  the proximity effect causes the LDOS in the metallic region to look
like that of a uniform $d$-wave superconductor, with peaks at nonzero energy. (See also Ref.~\onlinecite{nunner}.)

\section{Low order moments of the LDOS}

Moments of the one particle spectral function can be expressed in terms of matrix elements of the Hamiltonian and certain thermodynamic correlation functions, as has been recognized at least since the work of Nolting \emph{et al.}~\cite{Nolting}.  Nolting {\it et al.} and later authors, including Oganesyan~\cite{vadim}, Randeria \emph{et al.}~\cite{Randeria} and Norman 
\emph{et al.}~\cite{NormanARPES} proposed using a moment analysis to extract information from angle resolved photoemission (ARPES) data.
While the moment analysis can be carried through at a fundamental level, {\it i.e.} for Schrodinger's equation, we are actually more interested in interpreting the results in terms of a low energy effective theory.  We therefore begin by deriving expressions for the first and second moments of the LDOS in the presence of disorder (which was not considered by Nolting {\it et al}), and for different assumptions concerning the structure of the effective Hamiltonian.
\footnote{ A somewhat related analysis was  proposed by Randeria \emph{et al.}~\cite{Mottness}, based on a ``Mottness'' map which, within a single-band Hubbard model, relates certain integrals of the LDOS to the expectation value of the local electronic density, $n(\rr)$.}

We define the moments of the LDOS $\rho_{\rr}(\omega)$ as
\begin{equation}
 M_n(\rr)= \frac{\int^{\infty}_{-\infty} \omega^n\, \rho_{\rr}(\omega)d\omega}{\int^{\infty}_{-\infty} \rho_{\rr}(\omega)d\omega}\ . 
\end{equation}
(Note that an STM experiment does not actually measure the LDOS due to the existence of unknown matrix elements. However, since we have  normalized the moments as indicated, the effect of any energy independent matrix element cancels in  $M_n$.)  By definition, $M_0(\rr)=1$. 
Expressions for the first and second moments in the two cases discussed here are derived in Appendix A.

{\bf Case I:  A quadratic (mean-field) effective Hamiltonian:}
In systems in which the relevant physics can be described in mean-field approximation, the low energy physics can be derived from an effective Hamiltonian which is quadratic in the fermionic field operators;  this is the case in which the moment analysis is most powerful, as the moments depend only on the Hamiltonian, and are independent of the associated thermodynamic correlations.  

Needless to say, this approach neglects much of the strong correlation physics.  We partially address this issue in Case II below.  For now let us consider
the most general mean-field Hamiltonian,
\begin{eqnarray}\label{mf-Hamiltonian}
H =-\frac{1}{2}\sum_{\rr,\rr^\prime} \left\{t_{\rr,\rr^\prime}\psi^\dag_{\rr}\psi^{}_{\rr^\prime}
+  \Delta_{\rr,\rr^\prime}\psi^\dag_{\rr}\psi^\dag_{\rr^\prime}+{\rm H.c.}\right\},
\end{eqnarray}
for which the moments are straightforwardly seen to be 
\begin{eqnarray}
M_1(\rr) &=&- t_{\rr,\rr}\equiv -\mu_{\rr}  \label{eqn:mf-M1} \\
&& \nonumber\\
M_2(\rr) &=& 
\sum_{\rr^\prime} t_{\rr,\rr^\prime}t_{\rr^\prime,\rr}+ \sum_{\rr^\prime} |\Delta_{\rr,\rr^\prime}|^2. \label{eqn:mf-M2}
\end{eqnarray}
Here the index $\rr$ labels the position and band index of a Wannier function, as well as the spin-polarization, $\mu_{\rr}$ is the local value of the electro-chemical potential, $t_{\rr,\rr^\prime}$
is the hopping parameter from $\rr$ to $\rr^\prime$, and $\Delta_{\rr,\rr^\prime}$ 
is the superconducting gap parameter.

{\bf Case II:  Coupling to fluctuating fields:}
In Case I, we assumed that  $\Delta$, $\mu$ and $t$ were simply numbers.
However, these parameters could have dramatically fluctuating pieces, which are either present due to phonon mediated deformations of the effective electronic Hamiltonian or are generated (via a Hubbard-Stratanovich transformation) from an underlying interacting Hamiltonian. 
As discussed at the end of Appendix A, all the expressions in Eq.~\eqref{eqn:mf-M1} and \eqref{eqn:mf-M2} are generalized in this case  by replacing the left-hand side of each expression by its thermal expectation value (signified by $\langle \cdots\rangle$), so in Eq.~\eqref{eqn:mf-M1}, $\mu_\rr \to \langle \mu_\rr\rangle$ and in Eq.~\eqref{eqn:mf-M2}, $t_{\rr,\rr^\prime}t_{\rr^\prime,\rr}\to \langle t_{\rr,\rr^\prime}t_{\rr^\prime,\rr} \rangle$ and $|\Delta_{\rr,\rr^\prime}|^2\to \langle \Delta_{\rr,\rr^\prime}\Delta^\dagger_{\rr^\prime,\rr} \rangle$.  For example, a fluctuating superconducting gap parameter with rms value $|\Delta|$ (above $T_c$) makes the same contribution to the second moment as a mean-field gap of magnitude $|\Delta|$ (in the superconducting state).

Henceforth, we will focus on Case I, with the understanding that the effective parameters that enter the Hamiltonian may inherit temperature and magnetic field dependences from the underlying thermodynamic state of the system.

\section{Partial Moments}\label{sec:PM}

The integrals which define the moments, $M_n(\rr)$ extend to infinite energies.  Not only does this make them unmeasurable quantities, even if we could measure them, we  
would not expect them to yield useful information.  An effective Hamiltonian provides a reasonable description of the complex microscopic problem, at best, over a finite range of ``low'' energies.
Given this, we need to consider the more complex issue of what information can be gleaned from the partial moments, in which the
integrals are cutoff at a finite energy $\Omega$:
\begin{eqnarray}
M_n(\rr;\Omega) &\equiv& \frac{\int^{\Omega}_{-\Omega} \omega^n \rho_{\rr}(\omega) d\omega}{\int^{\Omega}_{-\Omega} \rho_{\rr}(\omega) d\omega}.\\
\end{eqnarray}
Note that $M_n(\rr)=M_n(\rr;\infty)$.

There is no simple general expression for the partial moments.  However, there are circumstances in which important information can be obtained from them, anyway.  In this sense, the application of these ideas is analogous to the application of the $f$-sum rule to the analysis of optical conductivity data.   For instance, were we interested in the temperature ($T$) or the magnetic field ($B$) dependence of the effective Hamiltonian, we might look at the difference between the LDOS at $T=0$ and $B=0$ and its value at non-zero $T$ or $B$;  if, as is often the case, $\rho_\rr(\omega)$ is essentially  $T$ and $B$ independent above a characteristic energy, $\Omega$, then by analyzing the partial moments $M_n(\rr,\Omega)$, we can obtain information about the induced {\it changes} in the effective Hamiltonian.  

For the present purposes, we will focus on a different example, relevant to the analysis of the above mentioned STM data in BSCCO.  Suppose we are interested in obtaining information about the spatial variations of the effective Hamiltonian.  Again, assuming that for a range of energies $\omega > \Omega$, $\rho_\rr(\omega)$ is approximately $\rr$ independent, we can analyze the spatially varying part of the moments, 
$\delta M_n\!\equiv\!M_n(\rr,\Omega)-\overline{ M}_n(\Omega)$ where $\overline{ M}_n(\Omega)$ is the average of $M_n(\rr,\Omega)$ for all $\rr$, in terms of the spatially varying pieces of the effective Hamiltonian.  An added advantage of this approach, as we shall shortly see, is that different terms in the effective Hamiltonian produce changes in the moments on different energy scales.
 
{\bf An explicit model:}  To illustrate the various aspects of this approach, we will consider
a specific model problem, which we design with the STM studies of BSCCO in mind.  We suppose that the
electronic structure is governed by a 2D effective Hamiltonian, $H^{eff}$ on a square lattice of the
form given in Eq.~\eqref{mf-Hamiltonian}, with parameters taken to be representative\cite{NormanBand,DHLee}
of the band structure and pairing symmetry of BSCCO.  
We consider the case in which there is an inhomogeneous ``patch'' of size $ 6\times 6$ embedded in an
otherwise uniform infinite background. (While this choice of patch size is arbitrary, we were  
motivated by the experimental observation in BSCCO \cite{lang, kapitulnik_2, scalapino} of apparent  
patches of size around 3nm.)
For the spatially uniform piece of $H^{eff}$ we define $t_{\rr,\rr^\prime} \equiv t$ for $\rr$
and $\rr^\prime$ nearest-neighbor sites.  (The band structure retains up to fifth nearest neighbor
terms, with second nearest neighbor $t_{\rr,\rr^\prime} = -0.55\,t$, and other hopping matrix elements
smaller in magnitude by a factor of five or more.)     The chemical potential is $t_{\rr,\rr} = \mu = 0.44 t$, and 
$\Delta_{\rr,\rr^\prime} = \pm \Delta_0$ for $\rr$ and $\rr^\prime$ nearest-neighbor sites with the signs chosen consistent
with the $d$-wave symmetry of the order parameter.  We further take a representative value of $\Delta_0 = 0.085\,t$.
(In the case of BSCCO, $t\approx 0.15$ eV and $\Delta_0 \approx 0.04$\,eV.)   
The band, which is not particle hole-symmetric, runs from $-2.7  t$ to $6.7  t$.

Inside the patch we will consider three different extreme cases: (1) Inhomogeneous electro-chemical
potential: $\mu_\rr = \mu + \delta \mu$ for $\rr$ inside the patch,  (2)  Inhomogeneous gap parameter with the $d-$wave symmetry:  $\Delta_{\rr,\rr^\prime} =  \Delta_0 \pm \delta\Delta$ for nearest neighbor $\rr, \rr'$ inside the patch.   
3)  Inhomogeneous band structure: 
$t _{\rr,\rr^\prime}= t + \delta t$ for nearest neighbor $\rr, \rr'$ inside the patch.

\begin{figure}[htbp]
	\centering
		\includegraphics[width=0.45\textwidth]{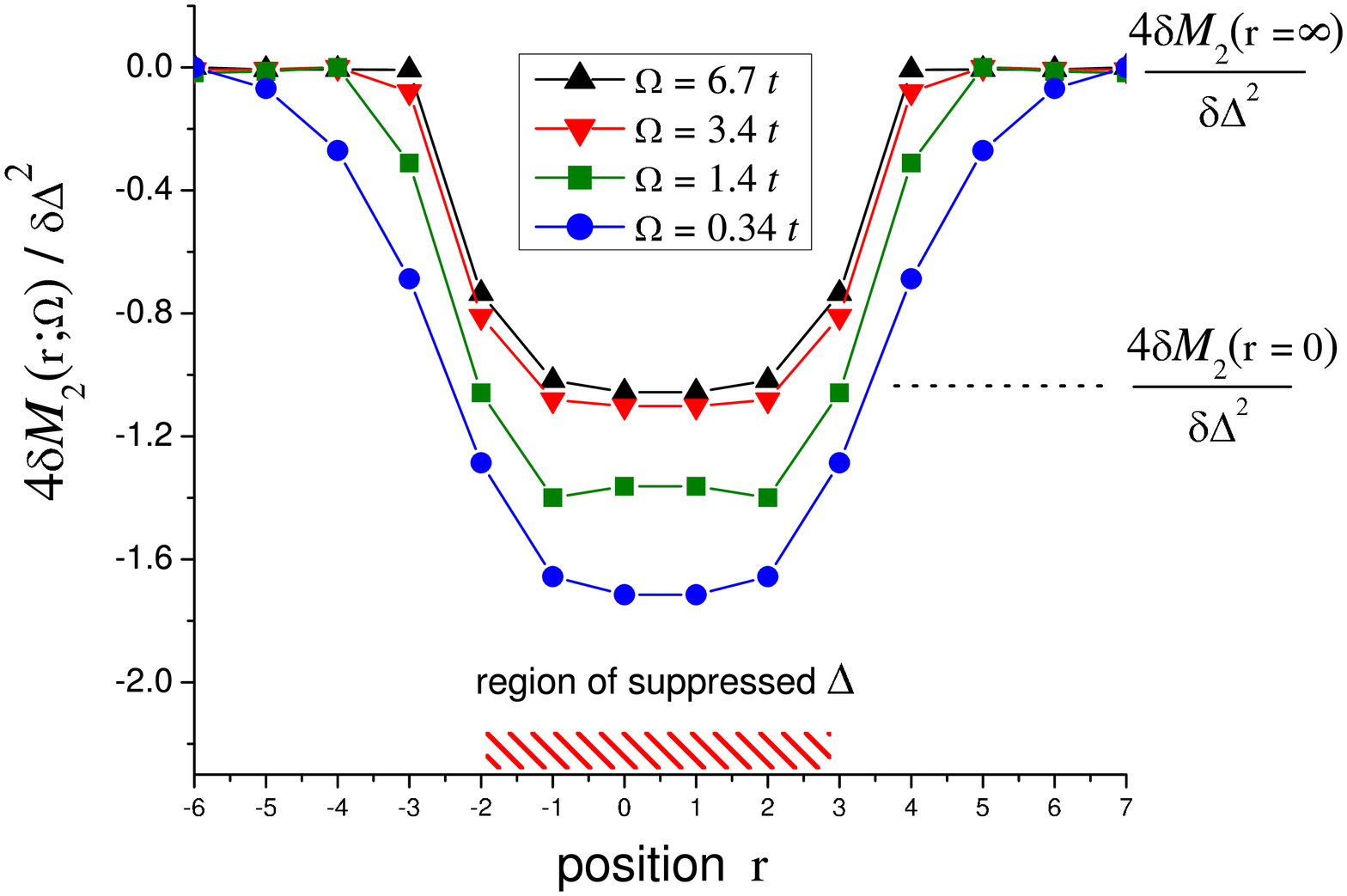}
	\caption{(Color online.)  The dimensionless response $\delta M_2(\rr;\Omega)/\delta\Delta^2$ due to the suppression of $\Delta$ in a $6\times 6$ patch (displayed region is 14 sites wide) 
	for several choices of $\Omega$.  (The band runs from $-2.7  t $ to $6.7  t$.)
	As we expect, when the cutoff is large, the second moment only knows about the local environment, and its
	value changes abruptly as $\rr$ crosses into the region of suppressed gap.
	Surprisingly, the response of $M_2(\rr;\Omega)$ to spatial
	variations of $\Delta$ remains quite local for $\Omega$ well above $\Delta_0\!=\!0.085t$.}\label{locality}
\end{figure}

An exact formal expression for the single-particle Green function of this problem can be obtained from multiple scattering
theory in terms of the Green function of the uniform system.  For each of the three cases, we have evaluated these expressions
numerically to compute the LDOS.  These are in turn integrated numerically to compute the partial moments.  

{\bf Locality of the partial moments:}
Fig.~\ref{locality} shows the response of ${M}_2(\rr;\Omega)$ to the inhomogeneity in the gap parameter for several choices
of $\Omega$. In the present case, we have taken $\delta \Delta = -\Delta_0$,
that is we have considered the case in which the pairing potential is absent inside the patch. For presentational clarity,
we have plotted $4\delta M_2(\rr;\Omega)/\delta\Delta^2$, the renormalized difference between $ M_2(\rr;\Omega)$ and the average value $\overline{M}_2(\Omega)$, which in this case is identical to the asymptotic value $\lim_{|\rr| \to \infty} M_2(\rr;\Omega)$). 
Hence $\delta M_2(\rr;\Omega)$ approaches 0 when the position $\rr$ is far from the patch.
Both the LDOS at fixed $\omega$ (not shown) and $M_2(\Omega)$ for $\Omega < \Delta_0$ exhibit complicated $\rr$ dependences,
with induced spatial variability that extends far outside the patch.  However, once $\Omega $ is greater than a few times $ \Delta_0$,  even though
$\Omega$ is still substantially less than the bandwidth, $M_2(\rr;\Omega)$ approaches its asymptotic value rapidly.   

From Eq.~\eqref{eqn:mf-M2} it follows that, in the limit $\Omega\!\to\!\infty$, 
$\delta M_2\!\to\!-\!\frac{\Delta_0^2}{4}\!=\!-\!\frac{(0.085\,t)^2}{4}\!\approx\!-\! 0.18\,t^2$.  For non-infinite $\Omega$, we can see that the
partial moment is already approximately equal to the $\Omega\!\to\!\infty$ moment when  $\Omega \approx 3\,t$ (\emph{i.e.}, half the bandwidth)
and it overestimates the response of the moment by $\approx$ 40\%  for $\Omega=1.4\,t$ (\emph{i.e.}, 1/5 of the bandwidth.) 
Thus, qualitatively accurate information about the structure of the effective Hamiltonian can
be obtained even if the range of integration is restricted to as little as 1/5 of the bandwidth.  Specifically, if one were to simply pretend
that $\delta M_2(\rr;\Omega=1.4 t)$ is approximately equal to $\delta M_2(\rr;\Omega=\infty)$, one would conclude that the gap
parameter changes from its value inside the patch to its value outside the patch over an interface region of width equal
to a couple of lattice constants (as opposed to the actual one lattice constant) and one would underestimate the magnitude
of the variation of the gap parameter by 40\%.\\
\begin{figure}[htbp]
\subfigure[]  {
\includegraphics[width=0.4\textwidth]{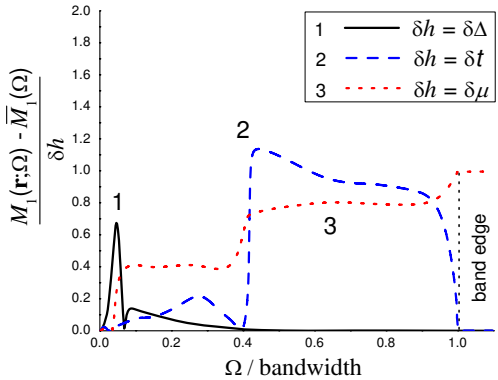}\label{omega1dependence}
}\\
\subfigure[]{
\includegraphics[width=0.4\textwidth]{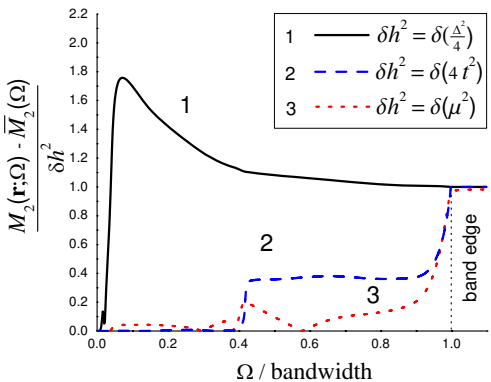}\label{omega2dependence}
}
\caption{(Color online.)  The difference between partial moments inside and outside the patch in response to the inhomogeneity in the parameter of the Hamiltonian $h=\Delta, \mu, t$ as a function of the cutoff $\Omega$. $\delta h=10$\,meV in all figures. 
	(a) The first moment difference (Eq.~\eqref{eqn:mf-M1}).	
	(b) The second moment difference (Eq.~\eqref{eqn:mf-M2}).
	}
\end{figure}

{\bf Distinguishing inhomogeneities in $\Delta$, $\mu$ and $t$:}
In Figs.~\ref{omega1dependence} and \ref{omega2dependence}  we have plotted the spatially varying part of the first and second moments, $\delta M_j(\rr,\Omega)= M_j(\rr,\Omega)-\overline{ M}_j(\Omega)$  for each of the three different types of inhomogeneity as a function of the cutoff, $\Omega$.  Here, $\rr $ is the point at the center of the patch.  Specifically, the three curves are for the case of  inhomogeneous electro-chemical
potential (dotted lines) with $\delta \mu = 0.1\, \mu$, inhomogeneous gap
parameter (solid lines)  with $\delta\Delta = - \Delta_0$ ({\it i.e.} vanishing gap parameter inside the patch) and  inhomogeneous band structure (dashed lines) with $\delta t = 0.1\, t$.  The curves have all been normalized, as indicated, to be appropriate dimensionless measures of the variation.  Since the spectrum is bounded above, the moments achieve their asymptotic values once $\Omega$ is in excess of the band edge, $ 6.7 t$.  The various kinks in the curves reflect the lower band edge.

In Fig.~\ref{omega1dependence} we see that in the case of an inhomogeneous $\mu(\rr)$, the first moment  has already reached 40\% of its asymptotic value by the time $\Omega$ is 1/20 of the bandwidth.  
The effect of the inhomogeneities in $\Delta$ and $t$ goes to zero for $\Omega$ in excess of the bandwidth as we expect from Eq.\eqref{eqn:mf-M1}. This happens very rapidly for the case of gap inhomogeneities, so even when $\Omega$ is moderately small compared to the bandwidth, the first moment gives a reasonable measure of the inhomogeneity of the band structure parameters, $\mu$ and $t$. 

The most striking feature of Fig.~\ref{omega2dependence} is that for small $\Omega$,  $\delta M_2(\rr;\Omega)$, becomes rapidly large (even larger than its asymptotic value) in the case of an inhomogeneous $\Delta$, but remains relatively small in the other two cases until $\Omega$ is a substantial fraction of the bandwidth.  (This is especially true for inhomogeneous $\mu$.)
We therefore conclude that $\delta M_2(\rr,\Omega)$  for $\Omega$ a few times $\Delta_0$ but still much less than the bandwidth gives a reasonable measure of the spatial variations of $\Delta$.

\subsection{LDOS near a unitary scatterer}
\begin{figure}[htbp]
\begin{minipage}[b]{.25\textwidth}
\subfigure[]{
\includegraphics[width=\textwidth]{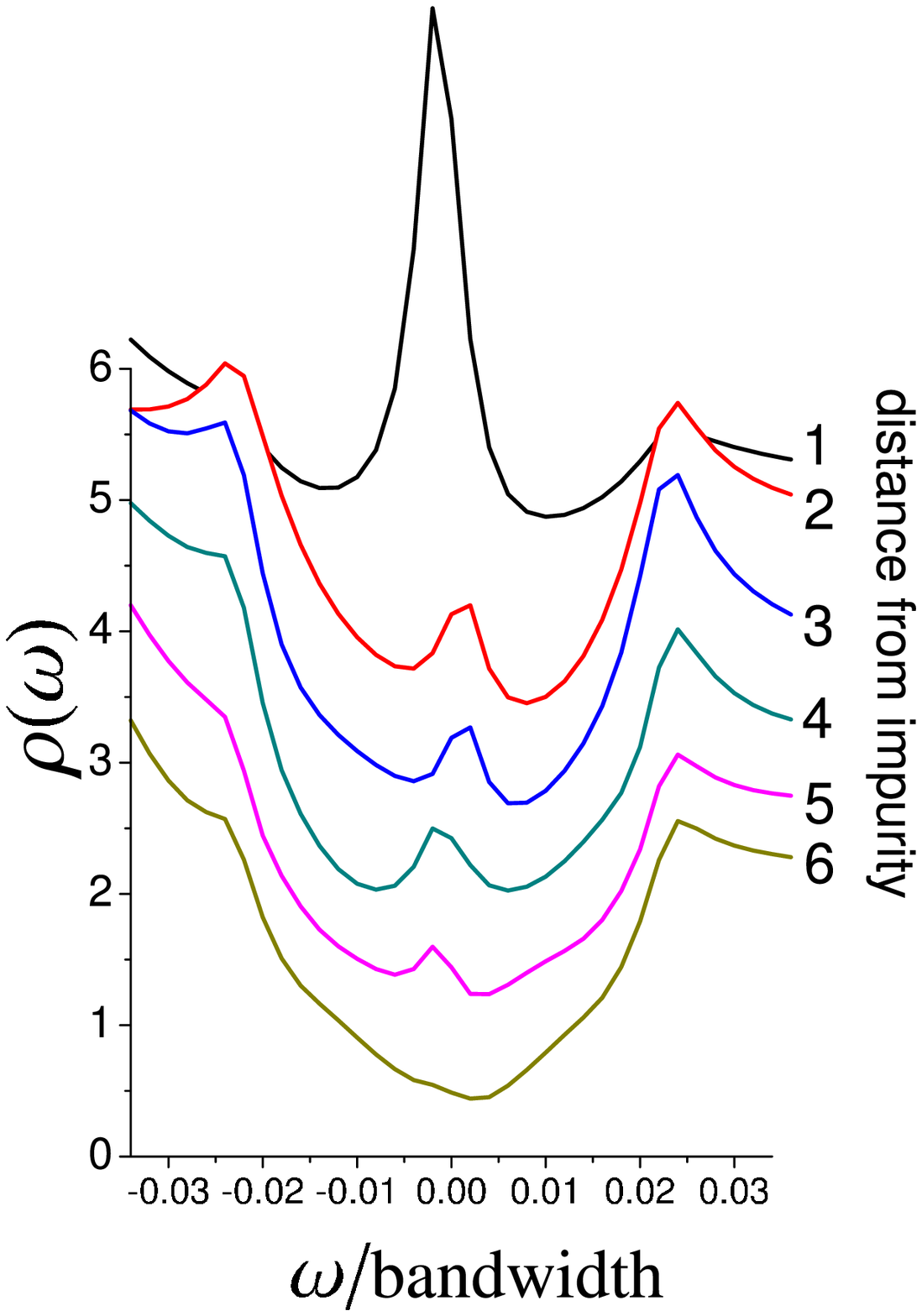}
}
\end{minipage}\hspace{4mm}
\begin{minipage}[b]{.18\textwidth}
\subfigure[]{
\includegraphics[width=\textwidth]{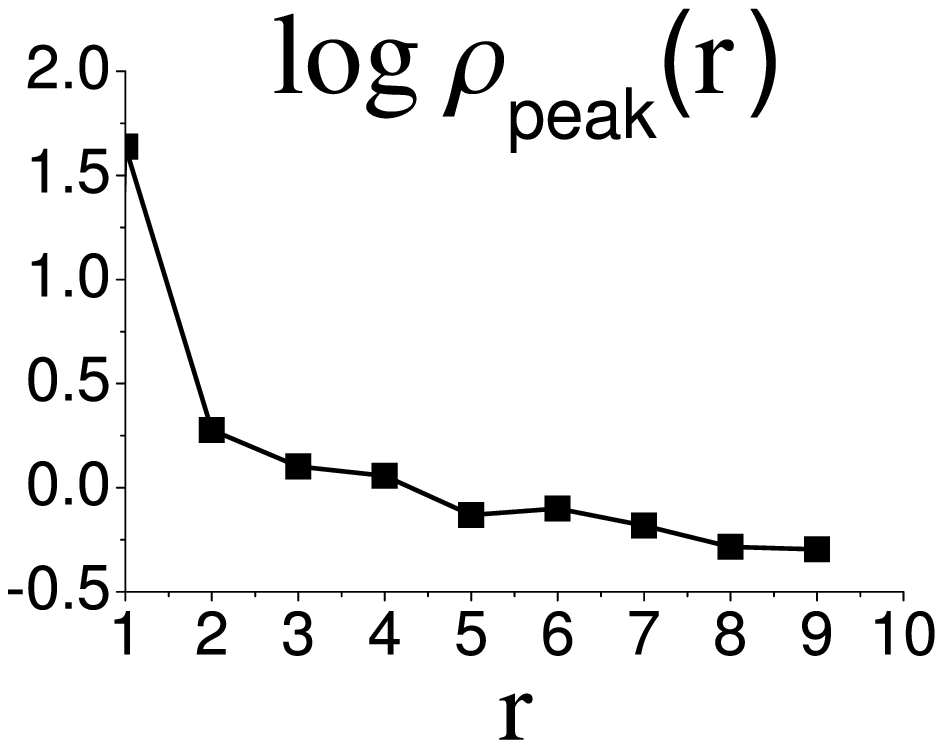}
}\\
\subfigure[]{
\includegraphics[width=\textwidth]{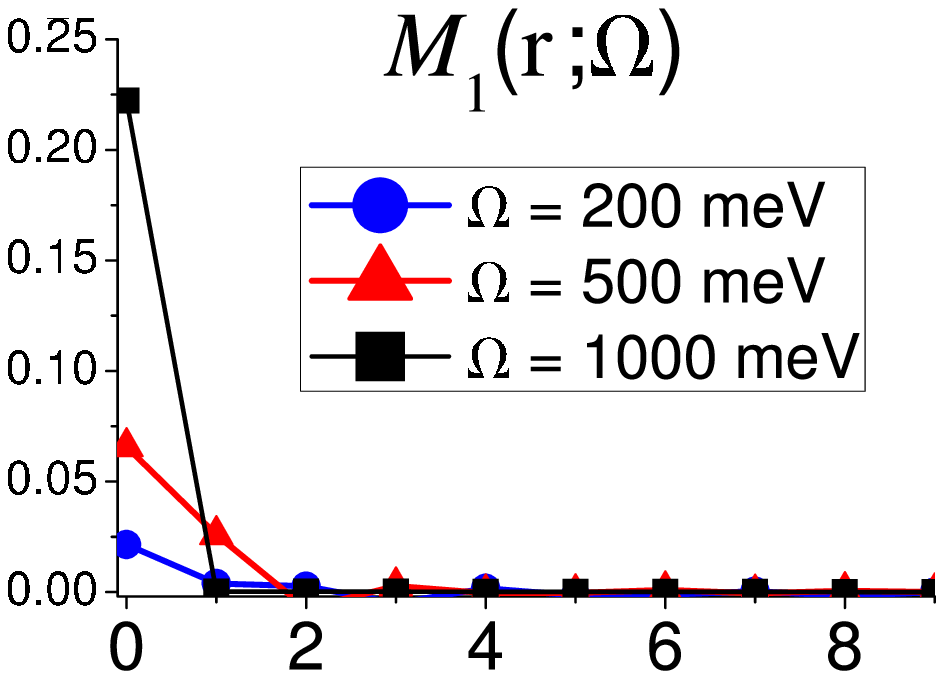}
}
\end{minipage}
\caption{\label{fig:LDOS_horizontal}(Color online.) 
(a) LDOS  $\rho(\omega)$ near a unitary scattering impurity for the model introduced in Sec.~\ref{sec:PM} at different positions.  
(b) Log of the peak height at $\omega\approx 0$ as a function of  distance from the impurity. 
(c) ${\delta M_1(\rr;\Omega)}$ along the bond direction for several choices of $\Omega$. 
}
\end{figure}

We have also analyzed an example of a single strong (unitary)
scatterer imbedded within an otherwise homogeneous
superconductor. 
The existence of a resonance near zero energy in the presence of  a unitary scatterer has been predicted by \textcite{balatsky_2} and experimentally confirmed \cite{davis_2, davis_3,yazdani_2}. 
Also finite energy resonances in the presence of a magnetic (non-unitary) impurity have been observed in Refs.~\cite{hudson,lang}

\begin{figure}[htbp]
	\centering
		\includegraphics[width=0.46\textwidth]{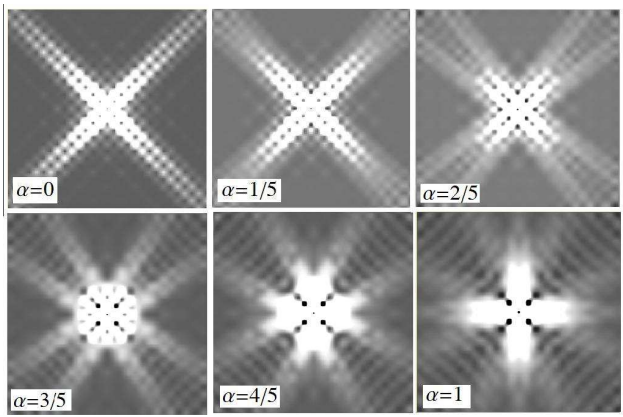}
	\caption{ LDOS at $\omega = 0$, near a unitary scatterer for several band structures,
$\epsilon($\textbf{k}$) = \epsilon_{nn}($\textbf{k}$) + \alpha (\epsilon_{0}($\textbf{k}$)- \epsilon_{nn}($\textbf{k}$))$ 
on a region of size $30\times 30$ sites. [Lighter shades denote larger $\rho($\textbf{r}$,\omega=0)$.]    
Here, $\alpha=0$ corresponds to nearest-neighbor tight-binding, whereas $\alpha=1$ corresponds to the full band structure from Refs. [\onlinecite{NormanBand,DHLee}]
used throughout this article. The short-range character of $\rho($\textbf{r}$,\omega=0)$ is sensitive to band structure, whereas the long-distance disturbance is strongest along, or near, the nodal direction.
}
\label{interp}
\end{figure}

Fig.~\ref{fig:LDOS_horizontal}(a) displays the LDOS $\rho(\omega)$ at different positions away from a unitary scatterer along the lattice direction (gap antinodal direction).  There is subtle fine structure in the LDOS which we have suppressed by introducing an energy broadening $\delta \omega =$2\,meV.  As is clear in the Fig.~\ref{fig:LDOS_horizontal}(a-b), there is a $\omega \approx 0$ resonance that peaks one site away from
the impurity ($\rho_{\rr=0}(\omega=0)=0$), and decays over the span of
several sites.  In the
direction $45^\circ$ to the nearest-neighbor bonds (nodal direction), the peak intensity is substantially
lower at short distances (see Fig.~\ref{interp}). 
As is seen in the Fig.~\ref{fig:LDOS_horizontal}(c), the first (partial) moment is much more highly localized,
even at relatively small $\Omega$, despite the fact that the impurity site
LDOS weight has been pushed almost entirely beyond the integration range.

As an aside, we note that it may be surprising that the disturbance caused by the impurity appears to extend more 
strongly in the antinodal direction than in the nodal direction, whereas it is theoretically clear that the asymptotic effect of the impurity 
at long distances is stronger in the nodal than in the bond direction.~\cite{balatsky}  
However, explicit calculations reveal that, to a remarkable extent, the structure at short and intermediate distances
depends strongly on the band structure.
This sensitivity to band structure is 
illustrated in Fig.~\ref{interp}: We have plotted the LDOS around a unitary scatterer, for a band structure $\epsilon(\kk) = \epsilon_{nn}(\kk)
+ \alpha (\epsilon_{0}(\kk)-\epsilon_{nn}(\kk))$, which interpolates between a simple nearest-neighbor band 
structure, $\epsilon_{nn}(\kk)$ (at $\alpha=0$), and the full band structure, $\epsilon_{0}(\kk)$ (at $\alpha=1$), used throughout 
this article.\footnote{The nearest-neighbor hopping in $\epsilon_{nn}(\kk)$ has been set larger than in $\epsilon_{0}(\kk)$ to insure the bandwidths 
are the same for both $\alpha = 0$ and $\alpha = 1$.}  As $\alpha$ increases from 0 to 1, the short range character changes from predominantly nodal to 
antinodal falloff at short distances.  Of course, the falloff is always slower in the nodal direction at long 
distances for any $\alpha \in [0,1]$.

%\begin{figure}[htbp]
%	\centering
%		\includegraphics[width=0.46\textwidth]{interp}
%	\caption{ LDOS at $\omega = 0$, near a unitary scatterer for several band structures,
%$\epsilon($\textbf{k}$) = \epsilon_{nn}($\textbf{k}$) + \alpha (\epsilon_{0}($\textbf{k}$)- \epsilon_{nn}($\textbf{k}$))$ 
%on a region of size $30\times 30$ sites. [Lighter shades denote larger $\rho($\textbf{r}$,\omega=0)$.]    
%Here, $\alpha=0$ corresponds to nearest-neighbor tight-binding, whereas $\alpha=1$ corresponds to the full band structure from Refs. [\onlinecite{NormanBand,DHLee}]
%used throughout this article. The short-range character of $\rho($\textbf{r}$,\omega=0)$ is sensitive to band structure, whereas the long-distance disturbance is strongest along, or near, the nodal direction.}
%\label{interp}
%\end{figure}

We have also looked at the case in which a resonant scatterer is placed at the center of a patch with either reduced or enhanced gap.  The behavior of the impurity induced disturbance of the LDOS is not qualitatively different in these cases, either - there is still a resonant peak near $\omega=0$ and its spatial decay occurs with a similar pattern.   Of course, other features of  the LDOS are even more complicated in these cases.
Nonetheless, even noting the strongly non-universal character of the induced features in the LDOS at short distances, the fact that
the low energy peak observed in STM studies of Zn doped BSCCO is sometimes seen to fall by as much as a factor of 100 in intensity 
within two or three lattice constants of a Zn impurity\cite{davis_2,davis_3} is not easily reconciled with a simple quasiparticle theory.

\section{BSCCO Analysis}
\begin{figure}[htbp]
\subfigure[]{
\includegraphics[height=0.142\textheight]{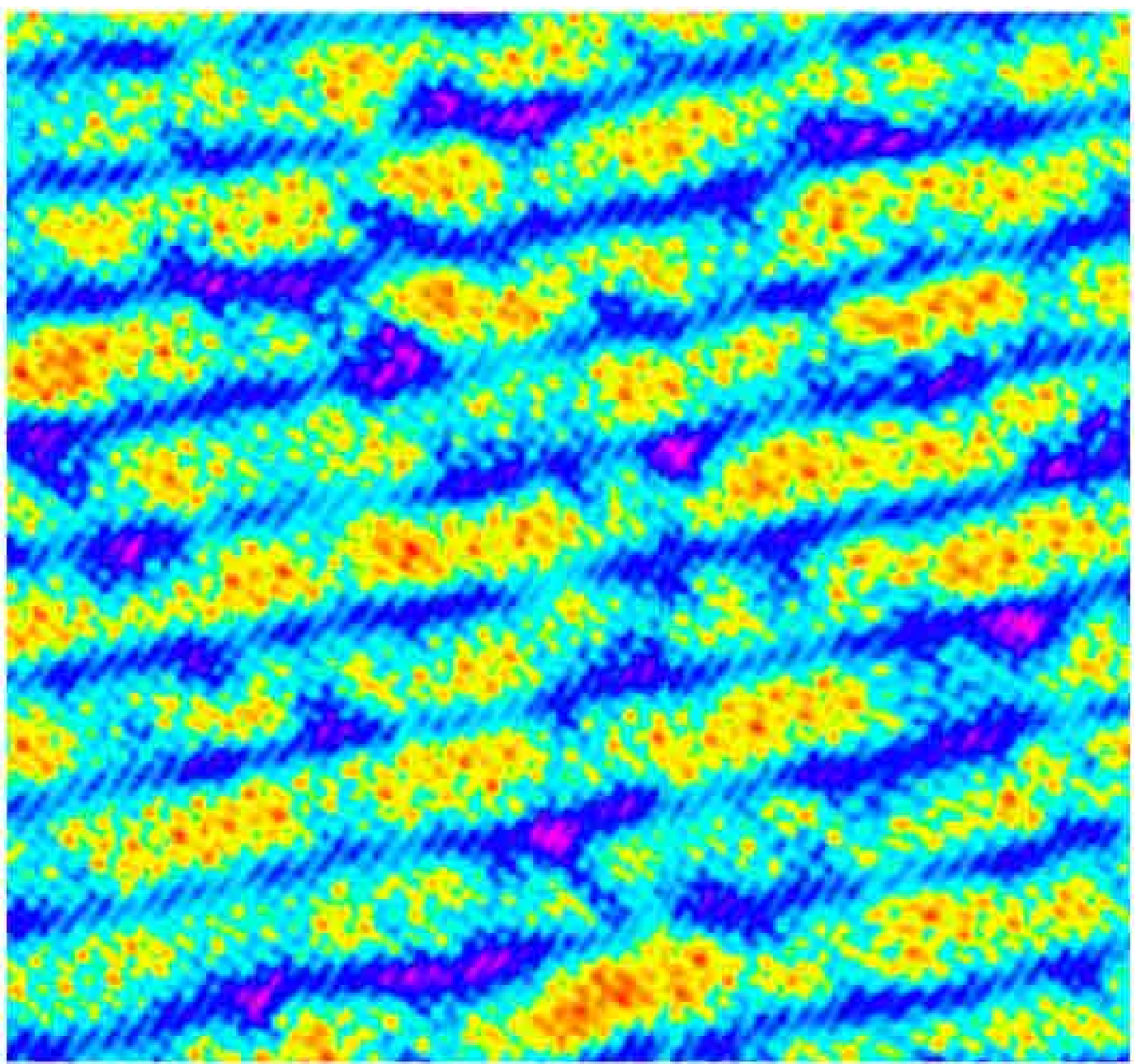}
}\hfill
\subfigure[]{
\includegraphics[height=0.142\textheight]{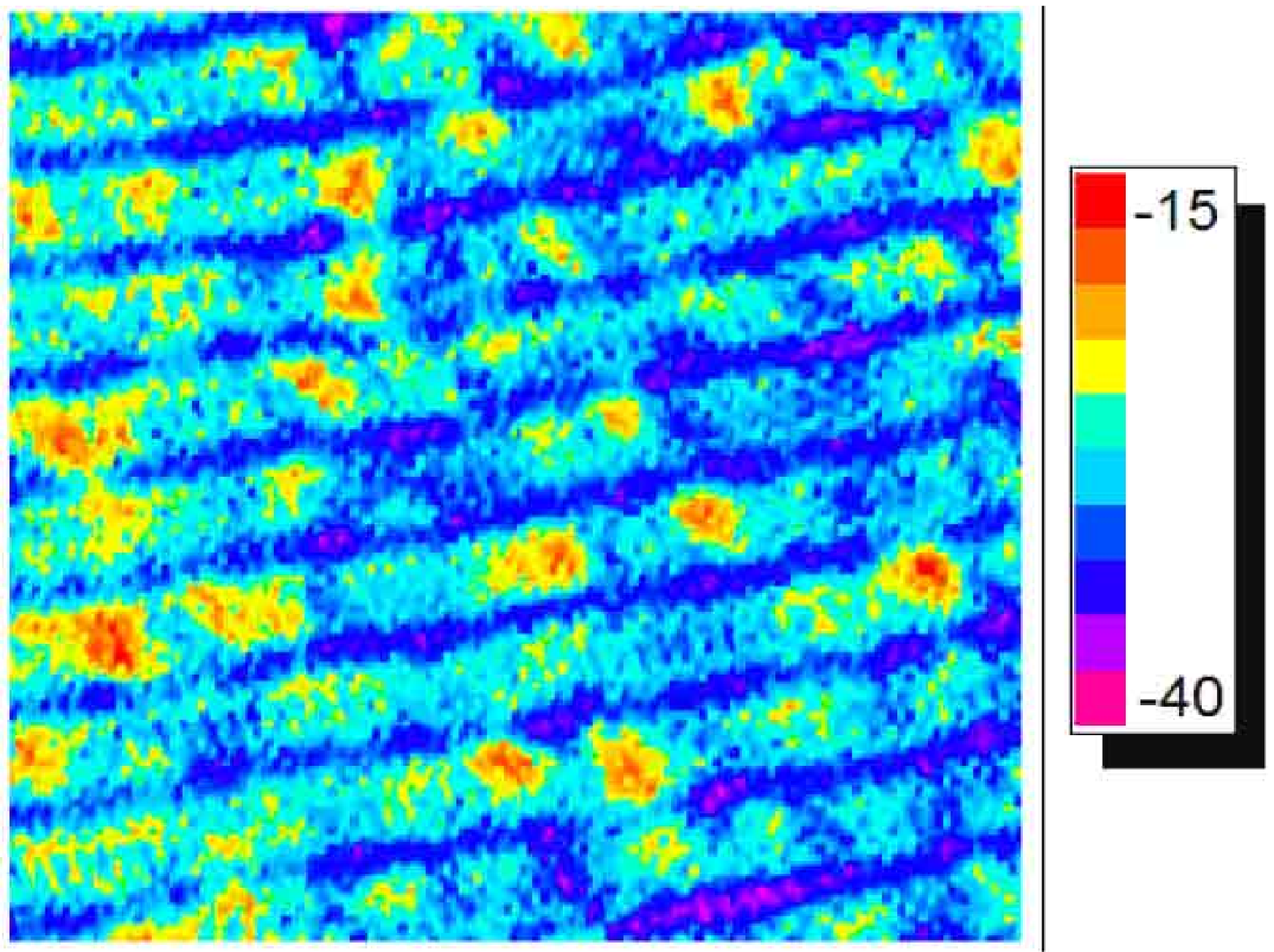}
}
\caption{\label{fig:moments1}
	(Color online.)
	(a) The topograph of BSCCO sample.
	(b) The first moment map $M_1(\rr;200\,\mathrm{meV})$ of the same sample. (Units in meV.) 
	The lattice superstructure is clearly visible in both.  
	}
\end{figure}

\begin{figure}[htbp]
\subfigure[]{
\includegraphics[height=0.17\textheight]{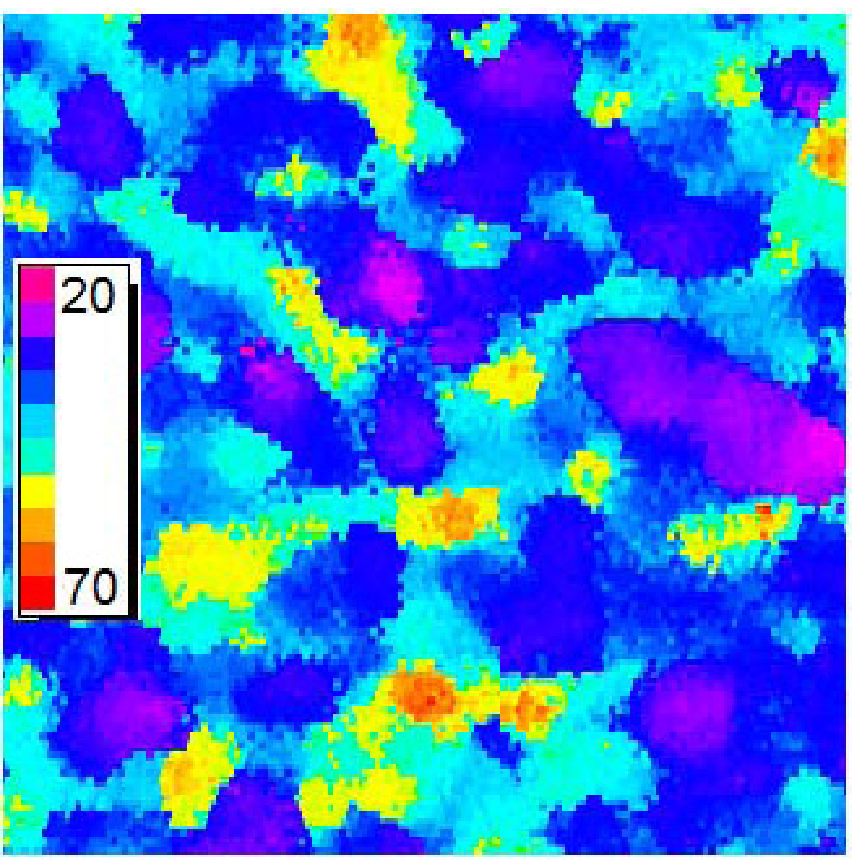}
}\hfill
\subfigure[]{
\includegraphics[height=0.17\textheight]{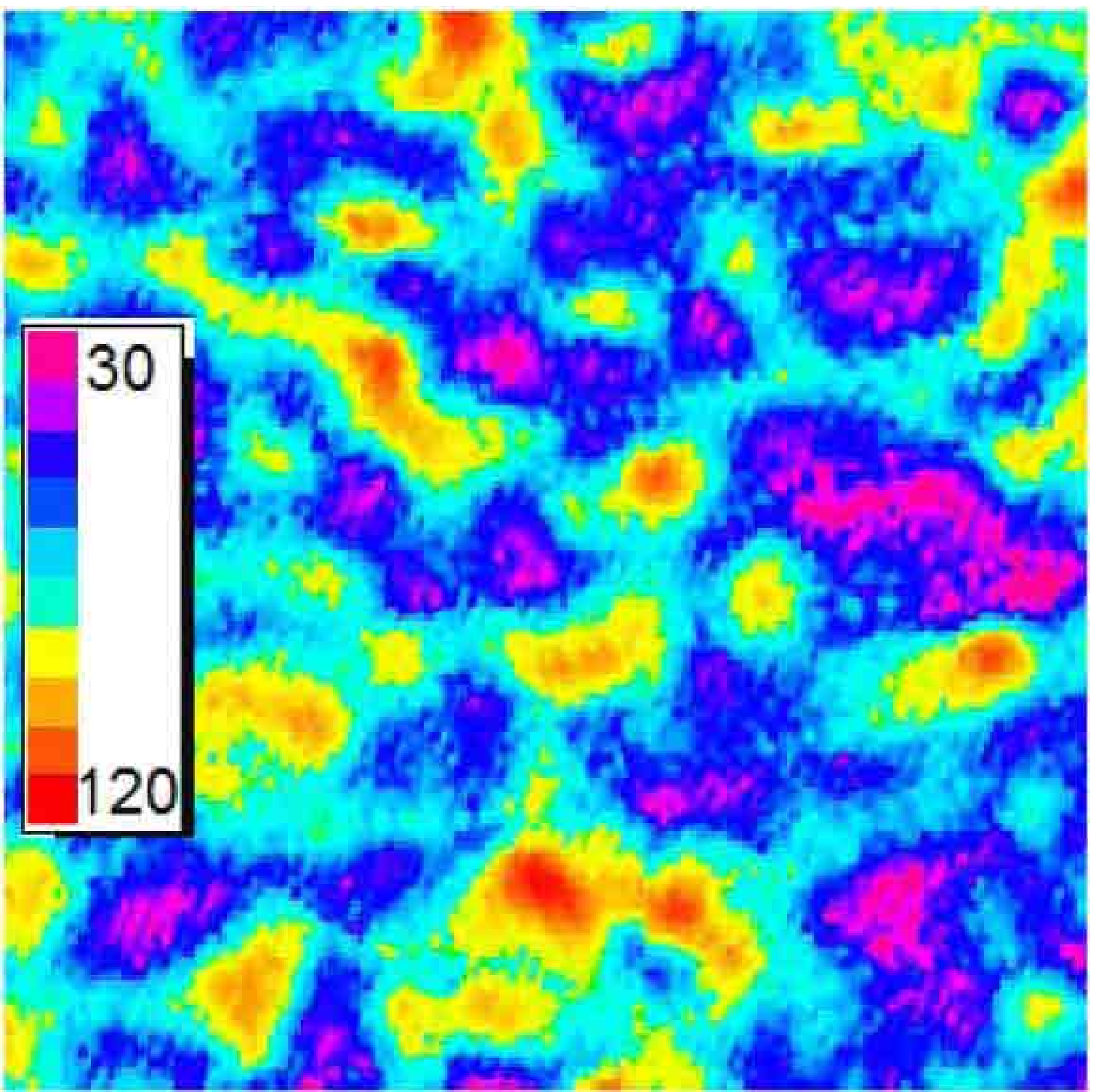}
}
\caption{\label{fig:moments2}
(Color online.)
(a) The gap map of BSSCO near optimal doping. (b) The map of the spatially varying piece of the gap magnitude from the second moment, $\sqrt{M_2(\rr;200\,\mathrm{meV})-M_2^{\textrm{min}}}$ from the same sample. (In determining $M_2^{min}$, the spatial minimum value of $M_2(\rr)$, we have ignored values of $M_2(\rr)$ that are smaller than the mean by more than three standard deviation.)}
\end{figure}

\begin{figure}[htbp]
{
\includegraphics[width=0.43\textwidth]{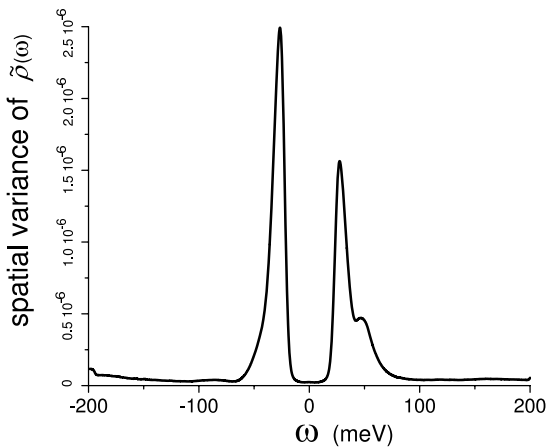}
}
\caption{\label{fig:variance}
The spatial variance of the normalized LDOS, $\tilde{\rho}(\omega)\!\equiv\!\frac{\rho(\omega)}{\int_{-\Omega}^{\Omega}d\omega\rho(\omega)}$ is calculated and plotted as a measure of degree of inhomogeneity of LDOS energy range between $-\Omega\!=\!-200\,\textrm{meV}$ and $\Omega\!=\!200\,\textrm{meV}$. As you can see the normalized LDOS is very homogeneous outside of the energy range between $-100\,\textrm{meV}$ and $100\,\textrm{meV}$. Also, as has been noted before, the LDOS is homogeneous at low energies as you can see in the flat region between $-25\,\textrm{meV}$ and $25\,\textrm{meV}$ and the surprisingly sharp onset of the inhomogeneity at about $\pm 25\,\textrm{meV}$.}
\end{figure}

As a test of the practical usefulness of the local moment analysis, we
have applied it to some STM data of Fang and
Kapitulnik\cite{scalapino} on a $220 \times 280\AA$ section of surface of near optimally doped {\BSCCO}.   
Fig.~\ref{fig:moments1} shows a plot of (a) the topograph and (b) the first moment $M_1(\rr;\Omega)$ with a cutoff  $\Omega = $200\,meV. (We chose this cutoff because it is several times the average gap maximum but still small compared to the bandwidth, and because at this energy the LDOS is more or less spatially uniform as Fig.~\ref{fig:variance} shows). Fig.~\ref{fig:moments2} shows a plot of (a) the gap
map and (b) the second moment also with $\Omega=$ 200\,meV.  (The gap map
displays the energy difference between the positive and the negative energy maxima in the LDOS at each point $\rr$.)
The superstructure due to the buckling of the Bi-O planes is clearly evident
in the topograph and the plot of $M_1$ (Fig.~\ref{fig:moments1}) but not in the gap map or the plot of the second moment (Fig.~\ref{fig:moments2}).   
Conversely, the splotchy character of the gap map, which has been interpreted as indicative of the existence of regions with different superconducting character, is largely reproduced in the second moment map, but seems largely uncorrelated with features in the topograph or first moment map.  

On the basis of the analysis of the previous section, we expect that $M_1(\rr, \Omega)$ is approximately proportional to the electro-chemical potential, even though $\Omega$ is small compared to the band width.  One of the most notable conclusions is that $M_1$ varies only by about $20\,\textrm{meV}$ from position to position.  Moreover, most of that variation 
is due to the superstructure variation.  Even if we  take into account the fact that
having a cutoff of $200\,\textrm{meV}$ may mean that the spatial variation in the electro-chemical potential, 
$\delta \mu$, may be as much as twice the variation in $M_1$, we still conclude that, in units of the band width (1.4 eV), the variations in the potential are small. 

The second moment varies as a function of $\rr$ by about $3000\,(\textrm{meV})^2$. On the basis of the results in Fig.~\ref{omega1dependence}, we feel it is safe to associate this with variations of $|\Delta|^2$.  However, because of the relatively small value of $\Omega$, direct application of Eq.~\ref{eqn:mf-M2} may overestimate the variations of $|\Delta|^2$ by as much as a factor of 2.

Assuming this factor of 2 overestimation
 i.e., $\delta M_2(\rr;200\,{\mathrm meV})\!\approx\!2\delta M_2(\rr) \approx 2\frac{\delta \Delta^2}{4}$ and that the regions with the smallest values of $M_2(\rr,\Omega)$ correspond to places where $\Delta(\rr)=0$, this analysis leads to the conclusion that
$\Delta$ varies as a function of position from  0 to about $80\,\textrm{meV}$; (Fang {\it et al.}\cite{scalapino} have shown the experimental data is compatible with minimum $\Delta = 0$.) Were we to assume a non-zero value for the minimum value of $\Delta$, we would obtain a correspondingly larger estimate of the maximum gap.  In comparison,  were we to interpret the  gap map as representative of $\Delta(\rr)$, the range of values inferred would be from approximately $25$ to $70$\,meV.  While both methods of analysis suggest large amplitude variations in $\Delta$, we believe that the present estimate, which implies even larger variations than does the gap map, is the more reliable.
This exaggerated response of the electronic correlations to
small variations in the local potential is suggestive of 
some form of frustrated phase separation\cite{vjeandsakphysicac}.

\subsection*{Acknowledgments}
We would like to thank A.V.Balatsky for helpful discussions. This work was supported through the National Science Foundation through grant
number NSF DMR 0531196 (S.K., J.R. and R.J.) and through the Department of Energy's Office of Science through grant No. DE-FG03-01ER45925 (A.F. and A.K.) and by the Stanford Institute for Theoretical Physics (E-A.K.).

\section{Appendix A}

Here we outline the proof for the sum rules, first for the mean field case.  The mean field Hamiltonian of Eq.~\ref{mf-Hamiltonian} can be written as 
\begin{equation}
H = \epsilon \sigma_3 + \Delta \sigma_1,
\end{equation}
in which $\epsilon$ and $\Delta$ are in general nontranslationally invariant hopping and gap
matrices, and $\sigma_j$ are the Pauli matrices with Nambu indices.  The Green function for this Hamiltonian can be defined as
\begin{equation}
G = \frac{1}{(\omega+i\delta)I-H}.
\end{equation}
In terms of $G$, the local density of states can be found as 
\begin{equation}\label{rho}
\rho_{\rr}(\omega) = -\frac{1}{\pi} \textrm{Im}\left[G_{1,1}(\rr,\rr)\right],
\end{equation}
where the subscripts are the Nambu indices.
We can decompose $G$ into the sum of a Hermitian and an anti-Hermitian operator. The Hermitian 
part obviously does not contribute to the LDOS. The anti-Hermitian part is equal to
\begin{equation}\label{g}
\frac{1}{2}(G-G^\dag)= -i\pi\delta(\omega I-H).
\end{equation}
Of course $\delta(\omega I-H)$ is defined as a matrix with the same eigenvectors $|\lambda\rangle$
as $H$, but with eigenvalues $\delta(\omega - \lambda)$. Eq.~\ref{rho} and \ref{g} tell us
\begin{equation}
\rho_{\rr}(\omega) =\left[\delta(\omega-H)\right]_{\rr 1;\rr 1}
\end{equation}
From this, we get the following relations:
\begin{eqnarray}
\int\omega^n\rho_{\rr}(\omega) d\omega&=& \int \omega^n\left[\delta(\omega-H)\right]_{\rr 1;\rr 1}d\omega \nonumber \\
                               &=& \left[\int\omega^n\delta(\omega-H)d\omega\right]_{\rr 1;\rr 1}  \nonumber \\
                               &=& H^n_{\rr 1;\rr 1}.
\end{eqnarray}
However,
\begin{eqnarray}
H^0_{\rr 1;\rr 1}&\equiv& 1_{\rr 1;\rr 1}=1 \\
H^1_{\rr 1;\rr 1}&\equiv& \epsilon_{\rr 1;\rr 1}=-\mu \\
H^2_{\rr 1;\rr 1}&\equiv& \left[\epsilon^2+\Delta^2\right]_{\rr 1;\rr 1}\\ \nonumber 
               &=& \mu_{\rr}^2 + \sum_{\rr^\prime} t_{\rr \rr^\prime}^2+ \sum_{\rr^\prime} \Delta_{\rr \rr^\prime}^2,\\ 
\end{eqnarray}
and this gives us the desired sum rules. 

For the general interacting Hamiltonian, we can get the
density of states from the imaginary part of the retarded Green's function:
\begin{eqnarray}
\rho_{\rr}(\omega) &=& -\frac{1}{\pi} \textrm{Im} \left[-i \int_0^\infty e^{i\omega t} \left\langle\left\{\psi^{\dag}_{\rr}(0),\psi_{\rr}(t)\right\}\right\rangle\right] \nonumber \\
&=& \frac{1}{2\pi} \int_{-\infty}^\infty e^{i\omega t} \left\langle  \left\{\psi^{\dag}_{\rr}(0),\psi_{\rr}(t)\right\}  \right\rangle d t,
\end{eqnarray}
and therefore
\begin{eqnarray} 
&&  \int^\infty_\infty \omega^n \rho_{\rr}(\omega) d\omega = \nonumber \\
&=& \frac{1}{2\pi}\int^\infty_\infty \left[\left(-i\frac{d }{d t}\right)^ne^{i\omega t}\right] \left\langle\left\{\psi^{}_{\rr}(t),\psi^{\dag}_{\rr}(0)\right\}\right\rangle d t\, d \omega   \nonumber \\
&=&           \frac{1}{2\pi}\int^\infty_\infty e^{i\omega t} \left\langle\left\{\left[\left(i\frac{d }{d t}\right)^n\psi^{}_{\rr}(t)\right],\psi^{\dag}_{\rr}(0)\right\}\right\rangle d t\, d \omega     \nonumber \\
&=&           \int^\infty_\infty \delta(t) \left\langle\left\{\left[\left(i\frac{d }{d t}\right)^n\psi^{}_{\rr}(t)\right],\psi^{\dag}_{\rr}(0)\right\}\right\rangle d t  \nonumber \\
&=&           \left\langle\left\{\left[\left[\psi^{}_{\rr},H\right],\ldots,H\right],\psi^{\dag}_{\rr}\right\}\right\rangle,
\end{eqnarray}
from which the most general counterparts of Eq.~\eqref{eqn:mf-M1}and \eqref{eqn:mf-M2} follow. 
As an application of this formula we can see that if in the Hamiltonian ~\eqref{mf-Hamiltonian}, $t_{\rr,\rr^\prime}$ and $\Delta_{\rr,\rr^\prime}$ are turned into dynamical fields by adding a term to the Hamiltonian that is a function of these fields and their momenta then Eq.~\eqref{eqn:mf-M1} should be replaced by
\begin{eqnarray}
M_1(\rr) &=& -\langle \mu_{\rr} \rangle 
\end{eqnarray}
in which $\langle \mu_{\rr} \rangle $ is the thermal average of the dynamical field $\mu_{\rr}$. For the second moment, Eq.\eqref{eqn:mf-M2} turns into 
\begin{eqnarray}
M_2(\rr) &=&  \sum_{\rr^\prime} \langle t_{\rr,\rr^\prime}t_{\rr^\prime,\rr}\rangle+ \sum_{\rr^\prime} \langle \Delta_{\rr,\rr^\prime}\Delta^{\dag}_{\rr^\prime,\rr}\rangle  \nonumber \\ 
&+ &  \langle [t_{\rr,\rr},H] \rangle. \label{last}
\end{eqnarray}
The expectation value of the commutator of the Hamiltonian with any operator in equilibrium is, of course, zero, so the last term drops out of Eq.\eqref{last}.

\end{document}